\documentclass[a4paper]{article}

\usepackage{INTERSPEECH2020}
\usepackage{xcolor}
\usepackage{cite}
\usepackage{multirow}

\title{SpEx+: A Complete Time Domain Speaker Extraction Network}
\name{Meng Ge$^{1,2}$, Chenglin Xu$^{2,4,*}$, Longbiao Wang$^{1, *}$, Eng Siong Chng$^2$, Jianwu Dang$^{1, 3}$, Haizhou Li$^{4, 5}$ \thanks{This work is supported by Human-Robot Interaction Phase 1 (Grant No. 192 25 00054), National Research Foundation (NRF) Singapore under the National Robotics Programme; AI Speech Lab (Award No. AISG-100E-2018-006), NRF Singapore under the AI Singapore Programme; Human Robot Collaborative AI for AME (Grant No. A18A2b0046), NRF Singapore; Neuromorphic Computing Programme (Grant No. A1687b0033), RIE 2020 AME Programmatic Grant; National Natural Science Foundation (61771333), Tianjin Municipal Science and Technology Project (18ZXZNGX00330). The work by H. Li is partly funded by the Deutsche Forschungsgemeinschaft (DFG, German Research Foundation) under Germany's Excellence Strategy (University Allowance, EXC 2077, University of Bremen, Germany). $^*$Corresponding author.}}
\address{
  $^1$ Tianjin Key Laboratory of Cognitive Computing and Application,\\ College of Intelligence and Computing, Tianjin University, Tianjin, China\\
  $^2$ School of Computer Science and Engineering, Nanyang Technological University, Singapore\\
  $^3$ Japan Advanced Institute of Science and Technology, Ishikawa, Japan\\
  $^4$ Department of Electrical and Computer Engineering, National University of Singapore, Singapore\\
  $^5$ Machine Listening Lab, University of Bremen, Germany}

\begin{document}

\maketitle
\begin{abstract}
Speaker extraction aims to extract the target speech signal from a multi-talker environment given a target speaker's reference speech.
We recently proposed a time-domain solution, SpEx, that avoids the phase estimation in frequency-domain approaches. Unfortunately, SpEx is not fully a time-domain solution since it performs time-domain speech encoding for speaker extraction, while taking frequency-domain speaker embedding as the reference. The size of the analysis window for time-domain and the size for frequency-domain input are also different. Such mismatch has an adverse effect on the system performance. To eliminate such mismatch, we propose a complete time-domain speaker extraction solution, that is called SpEx+.
Specifically, we tie the weights of two identical speech encoder networks, one for the encoder-extractor-decoder pipeline, another as part of the speaker encoder. Experiments show that the SpEx+ achieves 0.8dB and 2.1dB SDR improvement over the state-of-the-art SpEx baseline, under different and same gender conditions on WSJ0-2mix-extr database respectively. 

\end{abstract}

\section{Introduction}

In real-world speech communication, speech signal is often corrupted by background noise or speaker interference. It is desirable to have a front-end speech processing module to remove the background noise or to extract the foreground speech, such as, speaker verification \cite{rao2019target}, speech recognition \cite{zmolikova2017speaker,delcroix2018single}.

Speaker-independent blind source/speech separation (BSS) studies have seen major progress, such as DPCL  \cite{hershey2016deep,isik2016single,wang2018alternative}, DANet \cite{chen2017deep,luo2018speaker}, PIT \cite{yu2017permutation,kolbaek2017multitalker,xu2018single}, and TasNet \cite{luo2018real,luo2019conv}. It aims to separate each source from a speech mixture of a known number of speakers. Speech separation has to channel the right speaker to the right output voice stream, that is called the global permutation ambiguity challenge \cite{spex2020}. Speaker extraction takes a different strategy. It only extracts the target speaker's voice given a reference speech from target speaker, thus avoids the problem of global permutation ambiguity and doesn't require the knowledge about the number of speakers in the mixture. But it requires a reference speech.

A general approach for speaker extraction is conducted in frequency-domain, such as, SpeakerBeam \cite{vzmolikova2017learning,delcroix2018single,delcroix2019compact,vzmolikova2019speakerbeam}, Voicefilter \cite{wang2018voicefilter}, and SBF-MTSAL-Concat \cite{xu2019optimization}. The frequency-domain  method inherently suffers from a phase estimation issue that is required during the signal reconstruction. 
To avoid phase estimation, we recently proposed a time-domain solution called SpEx \cite{spex2020} that exploited a MFCC-based speaker embedding obtained from the speaker encoder to form a top-down auditory attention to the target speaker. SpEx was claimed a time-domain solution because it takes the time-domain mixture speech as input and extracts the target speaker's voice through an encoder-extractor-decoder network. However, the top-down auditory attention in SpEx system is supervised by a speaker encoder, that takes spectral features MFCC as input. As a result, there is a potential mismatch of latent feature space between the speech encoder and speaker encoder, that limits the efficiency and effectiveness of SpEx system. 

To address the above mismatch problem in SpEx, we propose a complete time-domain speaker extraction framework, that is called SpEx+. We propose to share the same network structures and their weights between two speech encoders. By doing so, the mixture speech input, and the reference speech input are represented in an uniform latent feature space. 

This paper is organized as follows. In Section 2, we motivate and design the proposed SpEx+ architecture. In Section 3, we report the experiments. Section 4 concludes the study.

\section{SpEx+ Architecture}
\begin{figure}[t]
	\centering
	\includegraphics[width=0.9\linewidth]{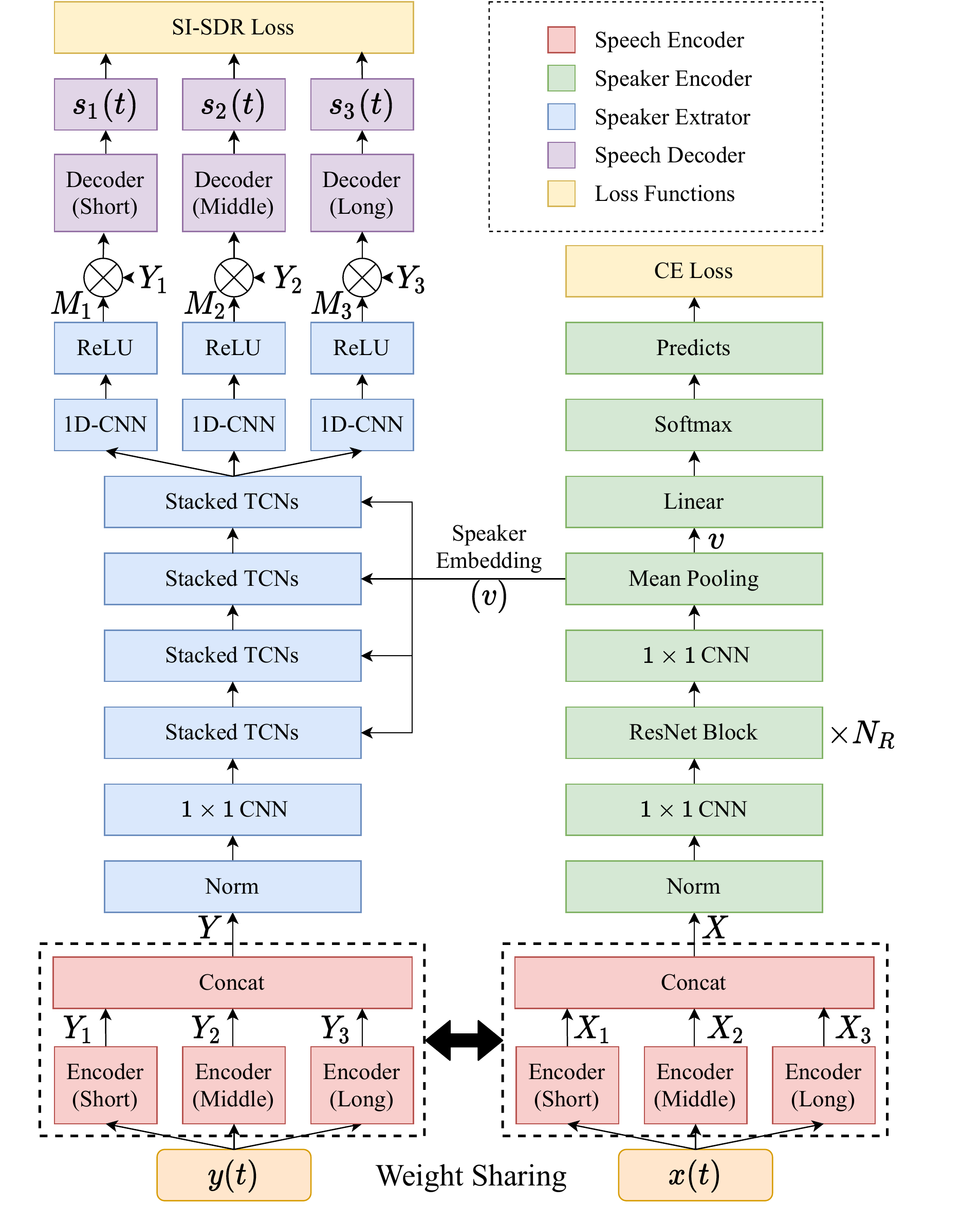}
	\caption{The diagram of the proposed SpEx+ system, which consists of two twin speech encoders, speaker encoder, speaker extractor, and speech decoder. ``Stacked TCNs'' represents a stack of several TCN blocks (i.e. 8 TCN blocks in this work) with exponential dilation increment. The details of TCN block and ResNet block are shown in Fig. \ref{fig:blocks}. The extracted signal $s_1(t)$ is chosen as the ultimate output at run-time inference.}
	\label{fig:speech_production}
\end{figure}

SpEx+ consists of speech encoder, speaker encoder, speaker extractor, and speech decoder as shown in Fig. \ref{fig:speech_production}, that has a similar architecture as SpEx \cite{spex2020}. The difference lies in the weight-shared speech encoders, also called twin speech encoders.


\subsection{Twin speech encoder}
Speech has a rich temporal structure over multiple time scales presenting phonemic, prosodic and linguistic content \cite{toledano2018multi}. It has shown that speech analysis of multiple temporal resolutions leads to improved speech recognition performance \cite{teng2016testing}. As shown in Fig. 1, we implement multi-scale speech encoding in speech encoder.

The speech encoder projects the input speech, either the mixture speech $y(t)$ to be extracted or the reference speech $x(t)$ from target speaker, into a common latent space. In SpEx+, the speech encoders support two processes, namely speaker encoding, and speech extraction. As the speaker encoding serves as the top-down voluntary focus for the speech extraction, we believe that it is beneficial for speaker encoding and speech extraction to share the same latent feature space. We propose a weight sharing strategy between the two speech encoders that have identical network structure as illustrated in Fig. 1. The two speech encoders run separately because they have to process different speech inputs. Therefore, we call them the twin speech encoders.

The twin speech encoders consist of several parallel 1-D CNNs with different filter lengths that result in various temporal resolutions. Although the number of multiple scales can be vary, we only study three different scales. The multi-scale embedding coefficients ($Y$ and $X$) of the mixture and the reference speech are then obtained from the speech encoders followed by a rectified linear unit (ReLU) activation function as follows:
\begin{align}
    Y &= [Y_1, Y_2, Y_3] = e(\theta|y,L_1,L_2,L_3,N) \\
    X &= [X_1, X_2, X_3] = e(\theta|x,L_1,L_2,L_3,N)
\end{align}
where $e(\cdot)$ represents a twin speech encoder, and $\theta$ is the shared parameters. $L_1(short)$, $L_2(middle)$ and $L_3(long)$ are the filter length of each filter to capture different temporal resolution in the 1-D CNN. To concatenate the embedding coefficients across different scale,
we use a same stride, $L_1/2$, across different scales. $N$ is the number of filters in the 1-D CNN.

Due to small filter size in time domain encoder, the twin speech encoder in time-domain takes a smaller window  than that in short time Fourier transform (STFT). As a result, the speech encoder generates embedding coefficients at a frame rate that is needed by the speaker extractor, however, much higher than necessary for the speaker encoder. The high frame rate may lead to gradient vanishing problem with recurrent neural network (RNN). We thus stack several residual convolutional blocks in the Speaker Encoder structure, as illustrated in Fig. 1, to obtain a suitable speaker embedding from the reference speech. Meanwhile, each residual convolutional block applies a max-pooling to take out the silence.

\begin{figure}[t]
	\centering
	\includegraphics[width=0.9\linewidth]{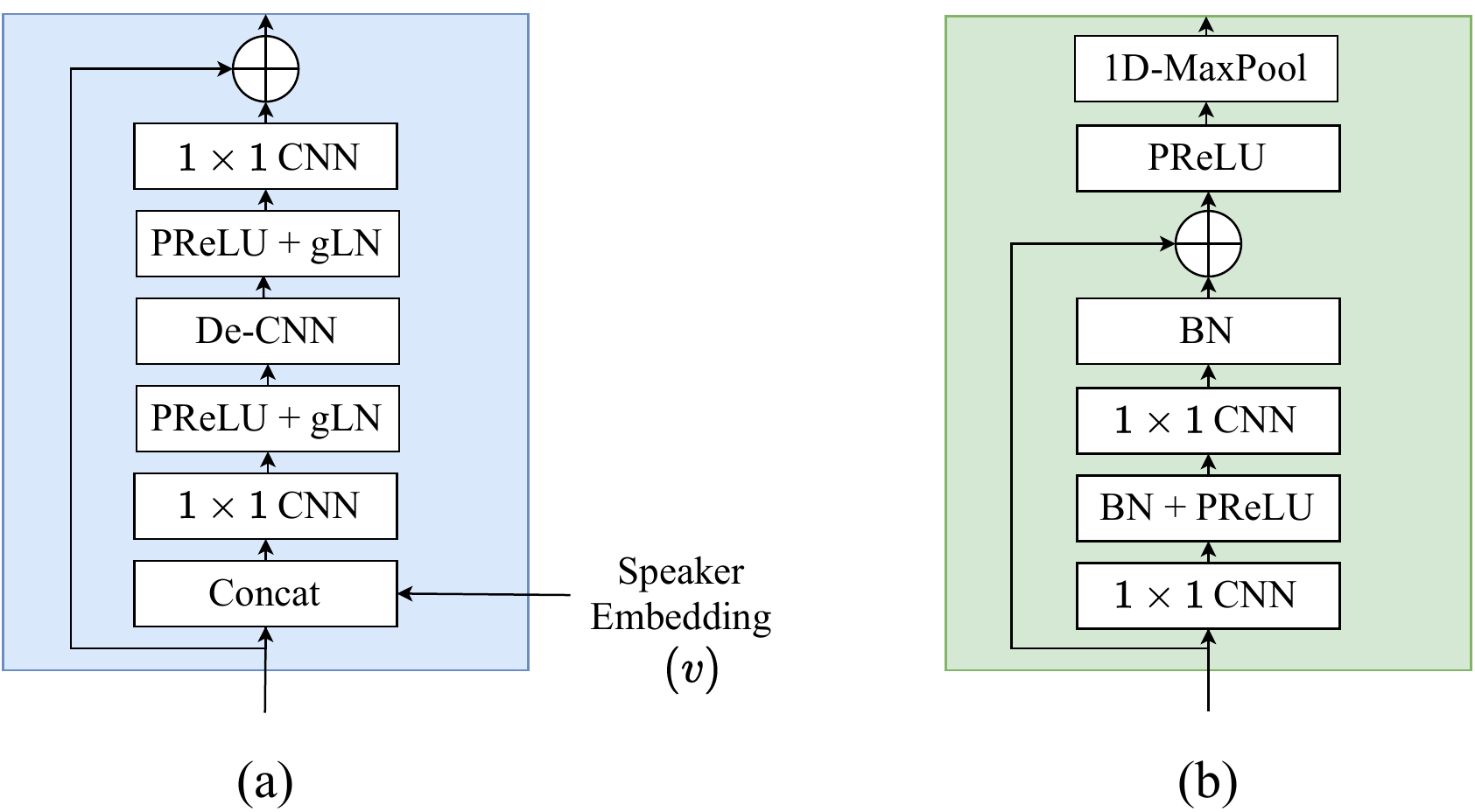}
	\caption{(a): The details of the first TCN block in each stack. Each stack has 8 TCN blocks in this work. The remaining 7 TCNs don't take the speaker embedding as additional inputs. ``gLN" is a global layer normalization. ``De-CNN" indicates a dilated depth-wise separable convolution. (b): The details of the ResNet Block. ``BN" is a batch normalization.}
	\label{fig:blocks}
\end{figure}

\subsection{Speaker encoder}

Speaker encoder is designed to extract speaker embedding of the target speaker from the reference speech. In practice, we employ a 1-D CNN on the embedding coefficients $X$ from the reference speech, followed by residual network (ResNet) blocks with a number of $N_R$. Then a 1-D CNN is used to project the representations into a fixed dimensional utterance-level speaker embedding $v$ together with a mean pooling operation. We have $v = g(X)$, where $g(\cdot)$ represents the speaker encoder. 

A ResNet block, as shown in Fig. \ref{fig:blocks}.b, consists of two CNNs with a kernel size of $1\times 1$ and a 1-D max-pooling layer. A batch normalization (BN) layer and parametric ReLU (PReLU) are used to normalize and non-linearly transform the outputs from each $1\times 1$ CNN. A skip connection is used to add the inputs to the activation representations from the second BN layer. The 1-D max-pooling layer with kernel size $1 \times 3$ tries to take out the silence. Meanwhile, the time series of the representations are reduced by 3 times.

The training of speaker encoder can be seen as a sub-task in a multi-task learning \cite{spex2020} of SpEx+ network. The speaker encoder can be jointly optimized by weighting a cross-entropy loss for speaker classification and a signal reconstruction loss \cite{wu2019time} for speaker extraction. During training, the gradients from two loss functions are back-propagated to optimize the speaker encoder for better speaker embedding. The details of the learning algorithm will be discussed in Section 2.4.

\subsection{Speaker extractor and speech decoder}

Speaker extractor is designed to estimate masks $M_i$ for the target speaker in each scale $i=1,2,3$, conditioned on the embedding coefficients $Y$ and the speaker embedding $v$. Similar to Conv-TasNet \cite{luo2019conv}, our speaker extractor repeats a stack of temporal convolutional network (TCN) blocks with a number of $B$ for $R$ times. In this work, we keep $B=8$ and $R=4$ same as in SpEx \cite{spex2020}. In each stack, the TCN has a exponential growth dilation factor $2^b (b \in \{0,\cdots,B-1\})$ in the dilated depth-wise separable convolution (De-CNN), as shown in Fig. \ref{fig:blocks}.a. The first TCN block in each stack takes the speaker embedding $v$ and the learned representations over the mixture speech. The speaker embedding $v$ is repeated to be concatenated to the feature dimension of the representation. 

Once the masks $M_i$ in various scales are estimated, the modulated responses $S_i=M_i \otimes Y_i$ are obtained by element-wise multiplication of the masks $M_i$ and the embedding coefficients $Y_i$. We then reconstruct the modulated responses $S_i$ into time-domain signals $s_i$ at multiple scales with the multi-scale speech decoder as follows:
\begin{align}
s_i &=d(M_i \otimes Y_i) \notag\\
     &=d(f(Y, v) \otimes Y_i)
\end{align}
where $\otimes$ is an operation for element-wise multiplication. $f(\cdot)$ and $d(\cdot)$ are the speaker extractor to estimate the mask and the speech decoder to reconstruct the signal, respectively.

\subsection{Multi-task learning}

The training of SpEx+ aims to achieve high quality speech, at the same time, highly discriminative speaker embedding for reference speech. Such strategy has proven to be beneficial in SpEx \cite{spex2020} and TD-SpkBeam \cite{delcroix2020improving}. We propose a multi-task learning implementation for SpEx+ training with two objectives, a multi-scale scale-invariant signal-to-distortion ratio (SI-SDR) loss for output speech quality, and a cross-entropy (CE) loss for speaker classification. The overall objective function is defined as,
\begin{equation}
\mathcal{L}(\Theta | y, x, s, I) = \mathcal{L}_{\text{SI-SDR}} + \gamma \mathcal{L}_{\text{CE}}
\label{eq2-1-4}
\end{equation}
where $\Theta$ represents the model parameters, $y$ is the input mixture speech, $x$ is the reference speech of the target speaker, $s$ is the target clean speech, $I$ is a one-hot vector representing the true class labels for the target speaker, and $\gamma$ is a scaling parameter.

$\mathcal{L}_{\text{SI-SDR}}$ aims to minimize the signal reconstruction error, which is defined as follows, with $\alpha$ and $\beta$ as the weights to different scales,
\begin{gather}
\mathcal{L}_{\text{SI-SDR}} = -[(1-\alpha-\beta)\rho(s_1, s) + \alpha \rho(s_2, s) + \beta \rho(s_3, s)], \notag\\
\rho(\hat{s}, s) = 20 \log_{10}\frac{||(\hat{s}^T s / s^T s) \cdot s||}{||(\hat{s}^T s / s^T s) \cdot s - \hat{s}||}
\label{eq2-1-5}
\end{gather}
where $\hat{s}$ and $s$ are the estimated signal and the target clean signal, respectively. Their means are normalized to zero. 

$\mathcal{L}_{\text{CE}}$ is the cross-entropy loss for speaker classification, which is defined as
\begin{equation}
\mathcal{L}_{\text{CE}} = -\sum\nolimits_{i=1}^{N_s} I_i \log(\sigma(W \cdot v)_i)
\end{equation}
where $N_s$ is the number of speakers in the speaker classification task, $I_i$ is the true class label for a speaker. $W$ represents a weight matrix, $\sigma(\cdot)$ is a softmax function and $\sigma(W \cdot v)$ represents the predicted probability.




\section{Experiments and Discussion}

\subsection{Dataset}

We simulated a two-speakers database WSJ0-2mix-extr\footnote{https://github.com/xuchenglin28/speaker\_extraction} at sampling rate of 8kHz based on WSJ0 corpus. The simulated database contains 101 speakers and was divided into three sets: training set (20,000 utterances), development set (5,000 utterances), and test set (3,000 utterances). Specifically, the utterances from two speakers in WSJ0 ``si\_tr\_s" corpus were randomly selected to generate the training and development set in a relative SNR between 0 to 5 dB. Similarly, the test set was generated by randomly mixing the utterances from two speakers in WSJ0 ``si\_dt\_05" and ``si\_et\_05" set. Since the speakers were unseen during training, the test set was considered as open condition evaluation. 

During the data simulation, the first selected speaker was chosen as the target speaker, the other one was regarded as the interference speaker. The required reference speech of the target speaker is a randomly selected utterance that is different from the utterance in the mixture. The reference speech is used to obtain a speaker embedding to characterize the target speaker.

\subsection{Experimental setup}

We trained all models for 100 epochs on 4-second long segments. The learning rate was initialized to $1e^{-3}$ and decays by 0.5 if the accuracy of validation set was not improved in 2 consecutive epochs. Early stopping was applied if no best model is found in the validation set for 6 consecutive epochs. Adam was used as the optimizer. 
The filter lengths of convolutions in speech encoder and decoder were $L_1=2.5$ms, $L_2=10$ms, $L_3=20$ms for speech of 8kHz sampling rate,
respectively. The speaker extractor follows the same configuration in SpEx\cite{spex2020}. The number of ResNet blocks in auxiliary network $N_R$ was set to 3, and the speaker embedding dimension was set to 256 in practice. For the loss configuration, we used $\alpha=0.1$, $\beta=0.1$, $\gamma=0.5$ to balance training loss. 

For TseNet \cite{xu2019time} baseline, 400-dim i-vector were extracted using Kaldi tool as speaker embedding. The speech encoder, speaker extractor and speech decoder are the same as SpEx+ except that single scale was used. For SpEx baseline system, we improved the performance from original 15.1 dB to 17.15 dB in terms of SDR by replacing the sigmoid activation function with ReLU function to estimate the masks. And we also retained the speech segments less than 4-seconds using padding operation, rather than discarded them as in the original SpEx system. The TseNet baseline was also re-implemented by adopting the above strategies. All systems were implemented with Pytorch\footnote{https://pytorch.org/}.

To examine the benefits of the twin speech encoders, we conducted two experiments, 1) SpEx+ (un-tied): the twin speech encoders share the same time-domain encoding network structure, but with independently trained network weights; 2) SpEx+ (tied): the twin speech encoders share the same time-domain encoding network structure, and the network weights are tied and updated together. We believe that SpEx+ (tied) configuration allows SpEx+ to project both the mixture speech and reference speech into the same latent space, that facilitates speech extraction process. 

\subsection{Comparative study on WSJ0-2mix-extr}

\renewcommand{\arraystretch}{1.5}
\begin{table}[tp]
	
	\centering
	\fontsize{7}{6.5}\selectfont
	\caption{SDR (dB), SI-SDR (dB) and PESQ of extracted speech for the proposed SpEx+ and other 3 competitive baseline systems under open condition.} 
	\label{tab:performance_comparison33}
	\begin{tabular}{|c|c|c|c|c|}
		\hline
		Methods&Domain&SDR&SI-SDR&PESQ\cr
		\hline
		\hline
		Mixture&- &2.60&2.50&2.31\cr\hline
		SpeakerBeam \cite{delcroix2018single}&Freq.&9.62 & 9.22 & 2.64 \\
		SBF-MTSAL-Concat \cite{xu2019optimization}&Freq. &11.39&10.60&2.77\cr
		TseNet \cite{xu2019time} &Time&15.24&14.73&3.14\cr
		SpEx \cite{spex2020}&Time&17.15&16.68&3.36\cr\hline
		SpEx+ (un-tied)  &Time&18.13&17.78&3.44 \cr
		SpEx+ (tied) &Time&\textbf{18.54}&\textbf{18.20}&\textbf{3.49}\cr
		\hline
	\end{tabular} \label{tbl:cmp_extr}
	\vspace{-5pt}
\end{table}

We compare the proposed SpEx+ network with other baseline systems in terms of SDR, SI-SDR and PESQ. From Table \ref{tbl:cmp_extr}, we conclude that: 1) SpEx+  significantly outperforms previous state-of-the-art TseNet and SpEx with relative improvements of 23.6\% and 9.1\% in terms of SI-SDR, respectively. The improvements mainly come from the multi-scale time-domain speaker encoder in the proposed SpEx+. The time-domain speaker embedding shows stronger ability in characterizing the target speaker than the i-vector and MFCC-based speaker embedding. SpEx+ could be regarded as a complete end-to-end solution as neither speaker characterization (i.e. i-vector) nor frequency-domain feature extraction (i.e. MFCC) is required externally. 2) By sharing a common speech encoder for between the mixture speech and the reference speech,  SpEx+ outperforms SpEx by using an unified latent feature space. 

\renewcommand{\arraystretch}{1.5}
\begin{table}[t]
	
	\centering
	\fontsize{7}{6.5}\selectfont
	\caption{SDR (dB) and PESQ in a comparative study of different and same gender mixture under open condition.}
	\label{tab:OC}
	\begin{tabular}{|c|c|c|c|c|c|c|}
		\hline
		\multirow{2}{*}{Methods}&
		\multicolumn{2}{c|}{SDR}&\multicolumn{2}{c|}{PESQ}\cr\cline{2-5}
		&Diff.&Same&Diff.&Same\cr
		\hline
		\hline
		Mixture&2.50 &2.70 &2.29 &2.34\cr\hline
		SpeakerBeam \cite{delcroix2018single} & 12.01 & 6.87 & 2.82 & 2.43 \\
		SBF-MTSAL-Concat \cite{xu2019optimization}&12.87 &8.84 &2.90 &2.54\cr
		TseNet \cite{xu2019time} &17.37 &12.80 &3.34&2.92\cr
		SpEx \cite{spex2020}&19.28 &14.72&3.53&3.16\cr\hline
		SpEx+ (un-tied) &19.58 &16.47&3.56 &3.30 \cr
		SpEx+ (tied)&\textbf{20.08} &\textbf{16.77}&\textbf{3.62}&\textbf{3.34}\cr
		\hline
	\end{tabular} \label{tbl:gender_cmp}
	\vspace{-10pt}
\end{table}

\begin{figure}[b]
	\vspace{-13pt}
	\begin{minipage}[t]{1\linewidth}
		\centering
		\includegraphics[width=1\linewidth]{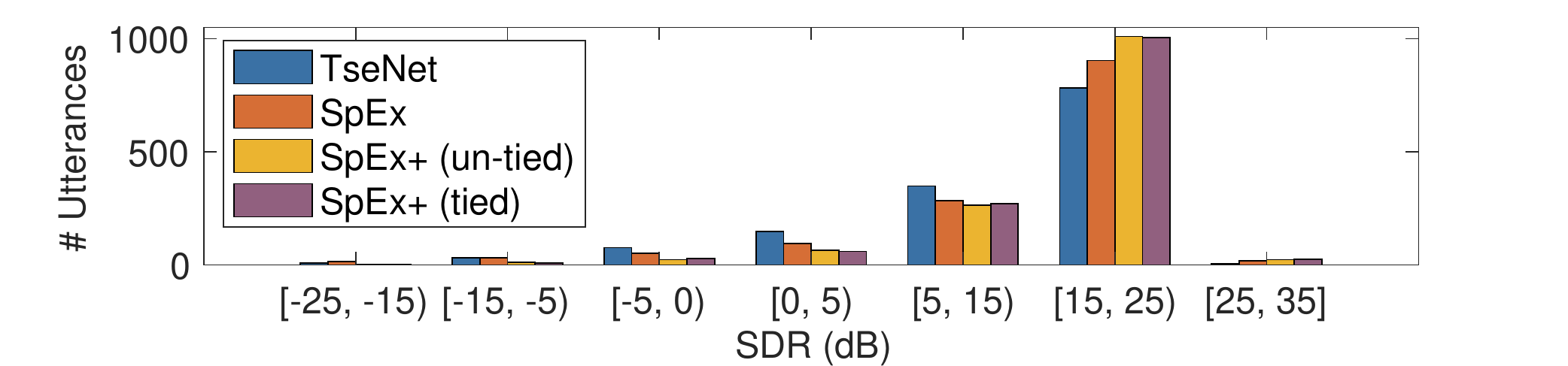}
	\end{minipage}%
	
	\begin{minipage}[t]{1\linewidth}
		\centering
		\includegraphics[width=1\linewidth]{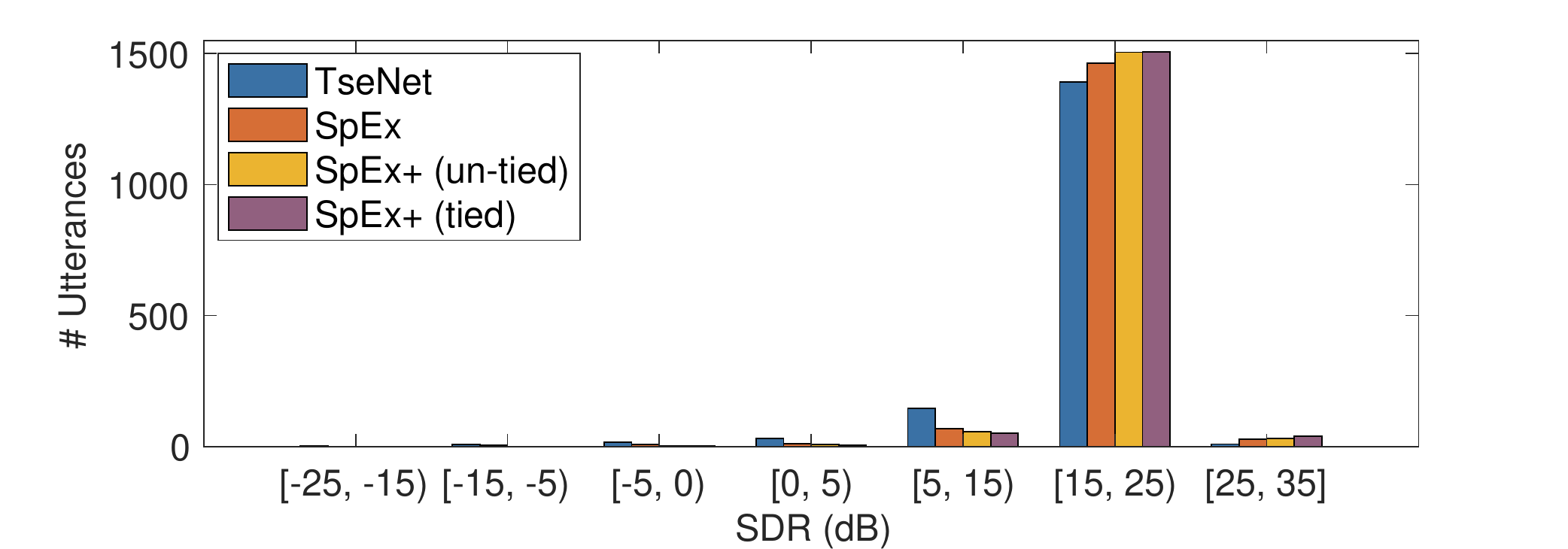}
	\end{minipage}
	\caption{Distributions of the number  of test utterances (open condition) at various dB ranges for same gender (top) and different gender (bottom) task. Higher SDR suggests better speaker extraction performance.}
	\label{fig:statistics}
\end{figure}

We further report the speaker extraction performance of various systems with different and same gender mixture speech, separately in Table \ref{tbl:gender_cmp}. We understand that same gender task is more challenging than different gender, thus, shows lower SDR and PESQ in general. It is observed that SpEx+ system outperforms all others in both different and same gender tasks. SpEx+ achieves 4.1\% and 2.5\% relative improvement over SpEx in different gender, and 13.9\% and 5.7\% relative improvement in same gender in terms of SDR and PESQ, respectively. Finally we report the SDR distributions of the 3,000 test utterances in Fig. \ref{fig:statistics}. We observe that the SpEx+ systems extract more utterances than other systems in the range of 15dB to 25dB.

\subsection{Comparative Study on WSJ0-2mix}
We first evaluate the effect of the duration of reference speech on WSJ0-2mix dataset, as reported in Table \ref{tbl:duration}. We observe that longer duration of reference speech always leads to better performance, that confirms our intuition.

\renewcommand{\arraystretch}{1.5}
\begin{table}[t]
	
	\centering
	\fontsize{7}{6}\selectfont
	\caption{SDRi (dB) and SI-SDR (dB) for different duration of reference speech on WSJ0-2mix. ``(avg.)" refers to the average duration of randomly chosen reference speech samples. The duration of the reference speech is randomly chosen during training in all experiments.}
	\label{tab:other_methods}
	\begin{tabular}{|c|c|c|c|}
		\hline
		Methods &Ref. Duration &SDRi &SI-SDR\cr
		\hline
		\hline
		\multirow{2}{*}{SpEx}&7.3s (avg.) &16.3 &15.8 \cr
		&60s &17.0 &16.6\cr\hline
		\multirow{5}{*}{SpEx+ (tied)}&7.3s (avg.) &17.2&16.9\cr
		&7.5s &17.2&16.9\cr
		&15s &17.5&17.2\cr
		&30s &17.6&17.3\cr
		&60s &\textbf{17.6} &\textbf{17.4} \cr
		\hline
	\end{tabular} \label{tbl:duration}
	\vspace{-5pt}
\end{table}

\renewcommand{\arraystretch}{1.5}
\begin{table}[t]
	
	\centering
	\fontsize{7}{6.5}\selectfont
	\caption{Speech separation and  speaker extraction on the WSJ0-2mix. For blind speech separation (BSS), we report the results evaluated on the oracle-selected streams. For speaker extraction (SE), we report the results on the extracted stream.}
	\label{tab:other_methods}
	\begin{tabular}{|l|l|c|c|c|c|c|}
		\hline
		Task&Methods&\#Params&SDRi&SI-SDR\cr
		\hline
		\hline
		\multirow{8}{*}{BSS}&DPCL++ \cite{isik2016single}&13.6M &-&10.8\cr
		&uPIT-BLSTM-ST \cite{kolbaek2017multitalker}&92.7M &10.0&-\cr
		&DANet \cite{chen2017deep}&9.1M&-&10.5\cr
		&cuPIT-Grid-RD \cite{xu2018single}&53.2M&10.2&-\cr
		&SDC-G-MTL \cite{xu2018shifted}&53.9M&10.5&-\cr
		&CBLDNN-GAT \cite{li2018cbldnn}&39.5M&11.0&-\cr
		&Chimera++ \cite{wang2018alternative}&32.9M&12.0&11.5\cr
		&WA-MISI-5 \cite{wang2018end}&32.9M&13.1&12.6\cr
		\hline
		BSS&BLSTM-TasNet \cite{luo2018real} &23.6M&13.6&13.2 \cr
		BSS&Conv-TasNet  \cite{luo2019conv}&5.1M&15.6&15.3 \cr
		BSS&DPRNN-TasNet \cite{luo2020dual} &2.6M&19.0&18.8\cr
		SE&SpEx \cite{spex2020}&10.8M&17.0&16.6 \cr
		\hline
		SE&SpEx+&11.1M&17.6&17.4\cr
		\hline
	\end{tabular} \label{tbl:cmp_separation}
	\vspace{-10pt}
\end{table}

We further compare a number of speaker extraction and speech separation techniques on WSJ0-2mix dataset. 
From Table \ref{tbl:cmp_separation}, we observe that SpEx+ achieves 13.7\% and 4.8\% relative improvements in terms of SI-SDR over the Conv-TasNet \cite{luo2019conv} for speech separation and SpEx \cite{spex2020} for speaker extraction, which employ same TCN blocks. We note that SpEx+ has the advantage over the group speech separation techniques (see BSS rows in Table 4) in dealing with unknown number of speakers, and global permutation ambiguity. A recent study shows that, by replacing the TCN block with a dual path RNN (DPRNN), DPRNN-TasNet \cite{luo2020dual} improves the performance of speech separation. We would explore the use of DPRNN in SpEx+ as our future work.

\section{Conclusions}

In this paper, we proposed a complete time-domain speaker extraction network. We proposed to share a multi-scale twin speech encoder to transform the mixture and reference speech into same latent feature space. We also proposed a multi-scale time-domain speaker encoder to obtain a speaker embedding that characterize the target speaker and guide the the speaker extraction from the mixture speech. Experiments showed that our proposed SpEx+ achieved significant performance improvement, especially in same gender mixture condition.

\bibliographystyle{IEEEtran}

\bibliography{IS2020}


\end{document}